\documentclass[12pt,preprint]{aastex}

\begin{document}

\title{Rapid Formation of Gas Giant Planets around M Dwarf Stars}

\author{Alan P.~Boss}
\affil{Department of Terrestrial Magnetism, Carnegie Institution of
Washington, 5241 Broad Branch Road, NW, Washington, DC 20015-1305}
\authoremail{boss@dtm.ciw.edu}

\begin{abstract}

 Extrasolar planet surveys have begun to detect gas giant planets
in orbit around M dwarf stars. While the frequency of gas giant
planets around M dwarfs so far appears to be lower than that around
G dwarfs, it is clearly not zero. Previous work has shown that
the core accretion mechanism does not 
seem to be able to form gas giant planets around M dwarfs, because the 
time required for core formation scales with the orbital period, which 
lengthens for lower mass stars, resulting in failed (gas-poor) cores 
unless the gaseous protoplanetary disk survives for $> 10$ Myr.
Disk instability, on the other hand, is rapid enough ($\sim 10^3$ yrs)
that it should be able to form gas giant protoplanets around even
low mass stars well before the gaseous disk disappears. A new suite 
of three dimensional radiative, gravitational hydrodynamical models 
is presented that calculates the evolution of initially marginally 
gravitationally unstable disks with masses of 0.021 to 0.065 $M_\odot$ 
orbiting around stars with masses of 0.1 and 0.5 $M_\odot$, respectively.
The models show that gas giant planets are indeed likely to form 
by the disk instability mechanism in orbit around M dwarf stars,
the opposite of the prediction for formation by the core accretion 
mechanism. This difference offers another observational test for
discriminating between these two theoretical end members for giant 
planet formation. Ongoing and future extrasolar planet searches 
around M dwarfs by spectroscopy, microlensing, photometry, and astrometry 
offer the opportunity to help decide between the dominance of the 
two mechanisms.

\end{abstract}

\keywords{stars: planetary systems -- stars: low-mass, brown dwarfs}

\section{Introduction}

 M dwarf stars dominate the nearby stellar population: within 10 pc of 
the sun, there are least 236 M dwarfs, but only 21 known G dwarfs
(Henry et al. 1997). In spite of this fact, spectroscopic extrasolar 
planet surveys have concentrated on G dwarfs, in hopes of finding Solar 
System analogues. Gravitational microlensing surveys at first set only upper 
bounds ($< 45\%$) on the frequency of multiple-Jupiter-mass planets orbiting 
between 1 AU and 7 AU around M dwarf stars toward the Galactic bulge 
(Gaudi et al. 2002), but recently the first microlensing detections 
of Jupiter-mass companions orbiting M dwarfs appear to have been 
accomplished (Bond et al. 2004; Udalski et al. 2005), and more should 
be on the way. Spectroscopic planet surveys have begun to focus more on 
M dwarfs, though a few M dwarfs have been included for several years. 
One of these, GJ 876, has a pair of gas giant planets (Marcy et al. 
1998, 2001) as well as an even lower mass planet of uncertain composition
(Rivera et al. 2005). Butler et al. (2004) and Bonfils et al. (2005a)
have found evidence for giant planets of Neptune-mass orbiting two
more M dwarfs, GJ 436 and Gl 581. There are hints 
from other spectroscopic surveys as well that M dwarfs may have
a significant frequency ($\le 13 \%$) of short period giant 
planets (Endl et al. 2003).

 Given that M dwarfs have masses as low as $\sim 0.1$ that of G dwarfs,
their most massive planets might be expected to be similarly lower
in mass and therefore harder to detect than those orbiting G dwarfs,
whose planets range up to masses of $\sim 10 M_J$ (Jupiter masses).
GJ 876 was the first M dwarf found spectroscopically to have planets, 
implying that its planetary spectroscopic signatures are relatively large. 
Hence, GJ 876's gas giants may be indicative of the most massive planets to 
be found around M dwarfs, at least on relatively short period orbits.
GJ 876's two gas giant planets have minimum masses of $\sim 2.1 M_J$ and
$\sim 0.56 M_J$ (Rivera et al. 2005), consistent with this argument,
given GJ 876's mass of $0.32 M_\odot$. If the range of gas giant masses
is shifted downward by a factor of $\sim 3$ or more for M dwarfs 
compared to G dwarfs, it will take considerably more effort to obtain
as complete a census of M dwarf planets compared to G dwarf planets, even
though the lower mass of the star itself works in favor of detection
by either spectroscopy or astrometry. The intrinsic faintness of M dwarfs
also adds to the observational burden, particularly for spectroscopic
surveys, which are photon-limited in their precision. 

 Young M dwarf stars appear to have dust disks similar to those around
T Tauri stars (Kalas, Liu, \& Matthews 2004; Liu et al. 2004; Liu 2004;
Mohanty, Jayawardhana, \& Basri 2005), so there is no {\it a priori} 
observational reason to believe that M dwarfs should not be able 
to form planets in much the same manner as their somewhat more 
massive siblings.

 Relatively little theoretical work has been done on the formation
of planets around low mass stars, because of the general obsession with
explaining the origin of our own planetary system. Wetherill (1996), 
however, found that Earth-like planets were just as likely to form 
from the collisional accumulation of planetesimals around M dwarfs 
with half the mass of the Sun as they were to form around G dwarfs.
Boss (1995) studied the thermodynamics of protoplanetary disks around 
stars with masses from 1.0 to 0.1 solar mass, and found that the location 
of the ice condensation point only moved inward by a few AU at most 
when the stellar mass was decreased through this range. In the core 
accretion model of giant planet formation (Mizuno 1980), this result 
implied that gas giant planets should be able to form equally well 
around M dwarfs, though perhaps at somewhat smaller orbital distances. 

 Core accretion, however, is a process once thought to have required times 
of order 10 Myr to form a core massive enough to accrete a gaseous envelope
orbiting a G dwarf star (Pollack et al. 1996). Theorists have worked hard to 
shorten the core accretion time scale to as little as $\sim 4$ Myr at 5.2 AU
(Inaba, Wetherill, \& Ikoma 2003), given the evidence for typical disk
lifetimes of $\sim 3$ Myr (Haisch, Lada, \& Lada 2001;
Eisner \& Carpenter 2003) or less in regions of high mass star 
formation (Bally et al. 1998; Bric\~eno et al. 2001). The time scale 
for core growth can be shortened to $\sim 1$ to 2 Myr by assuming 
that inward Type-I migration hastens the accumulation of solid cores 
(Alibert et al. 2005), but Type-I migration appears to be more
of a fatal danger to core accretion in the presence of gas
than an aid to gas giant planet formation (Kominami, Tanaka,
\& Ida 2005): in spite of efforts to lengthen them, Type-I migration 
time scale estimates remain considerably shorter than core growth
time scales and gaseous disk lifetimes, and considerably shorter 
than was assumed to be the case by Alibert et al. (2005). 

 In a turbulent, magnetorotationally unstable disk, Type-I migration 
can become at least in part a stochastic process, resulting in a 
component of random walk in semimajor axis and hence a distribution
of time scales for inward migration (Laughlin, Steinacker, \& 
Adams 2004; Nelson 2005), possibly allowing some cores to accrete
gaseous envelopes before migrating inward. However, such turbulence
is limited to ionized regions of the disk, whereas the planet-forming
midplane from $\sim$ 1 AU to $\sim$ 10 AU is likely be magnetically
dead (e.g., Gammie 1996), leaving Type-I migration as a significant
danger for core accretion.

 Core accretion is expected to take even longer to produce a
gas giant planet around an M dwarf, as the lower stellar mass
leads to longer orbital periods at a given distance from the star.
Core accretion thus appears to be too slow to produce Jupiter-mass 
planets in orbit around M dwarfs before the disk gas disappears 
(Laughlin et al. 2004). Laughlin et al. (2004) found that core
accretion required significantly more than $\sim 10$ Myr to form
a gas giant in orbit around a $0.4 M_\odot$ star, compared to
a similar calculation where a gas giant formed in $\sim 3$ Myr
around a $1 M_\odot$ star. Roberge et al. (2005) found no evidence 
for significant gas in the debris disk of the M1 star AU Microscopii, 
with an age of $\sim 12$ Myr. Laughlin et al. (2004) concluded
that gas giant planets should be rare around M dwarfs, but that
failed cores (i.e., roughly Neptune-mass cores that grew too slowly 
to accrete a significant gaseous envelope) should be common.

 Metallicities are notoriously difficult to measure in M dwarfs,
but a recent study of wide binary systems including M dwarf secondaries
by Bonfils et al. (2005b) has permitted a photometric calibration
for M dwarf metallicities, assuming that the M dwarf secondaries
have the same metallicities as their F, G, or K primaries. Bonfils 
et al. (2005b) thereby determined that the metallicities of two of the 
M dwarfs (Gl 876 and Gl 436) known from Doppler spectroscopy to have
giant planets are very close to solar. They also found that the
metallicities of 47 nearby M dwarfs are slightly lower on average
than those of nearby F, G, or K dwarfs, consistent with the M dwarfs
being somewhat older on average. These results suggest that M dwarfs
are able to form giant planets even with solar metallicities, though
the sample size obviously needs to be increased greatly.

 In this paper we examine the possibility of forming gas giant
planets around M dwarfs by the competing mechanism of disk instability 
(Cameron 1978; Boss 1997, 1998, 2003, 2004, 2005; Mayer et al. 
2002, 2004). In the disk instability mechanism, a marginally
gravitationally unstable protoplanetary disk forms spiral arms that can
lead to the formation of gravitationally-bound clumps of Jupiter-mass
on a time scale of $\sim 10$ orbital periods, typically $\sim 10^3$ yrs 
or less for a protoplanetary disk at orbital distances of $\sim 10$ AU. 
Disk instability represents a potentially rapid mechanism for gas giant
planet formation around M dwarfs, if M dwarfs have suitable marginally
gravitationally unstable disks. 

\section{Numerical Methods}

 The calculations were performed with a finite volume code
that solves the three dimensional equations of hydrodynamics and
radiative transfer, as well as the Poisson equation for the gravitational 
potential. The code is second-order-accurate in both space and time 
(Boss \& Myhill 1992) and has been used extensively in previous 
disk instability studies (e.g., Boss 2003, 2004, 2005). 

 The equations are solved on spherical coordinate grids with 
$N_r = 101$, $N_\theta = 23$ in $\pi/2 \ge \theta \ge 0$, 
and $N_\phi = 256$ or 512. The radial grid extends from 
4 AU to 20 AU with a uniform spacing of $\Delta r = 0.16$ AU.
The $\theta$ grid is compressed toward the midplane in order to ensure 
adequate vertical resolution ($\Delta \theta = 0.3^o$ at the midplane). 
The $\phi$ grid is uniformly spaced to prevent any azimuthal bias. 
The central protostar wobbles in response to the growth of 
disk nonaxisymmetry, preserving the location of the center 
of mass of the star and disk system. The number of terms in the 
spherical harmonic expansion for the gravitational potential of the disk
is $N_{Ylm} = 32$ or 48. The Jeans length criterion is monitored
throughout the calculations to ensure proper spatial resolution: the 
numerical grid spacings in all three coordinate directions should
remain less than 1/4 of the local Jeans length in order to avoid
the possibility of spurious fragmentation caused by poor spatial
resolution.
 
 The boundary conditions are chosen at both 4 AU and 20 AU to absorb radial 
velocity perturbations. Mass and linear or angular momentum entering 
the innermost shell of cells at 4 AU is added to the central protostar
and thereby removed from the hydrodynamical grid. During a typical
model, several Jupiter-masses of disk gas are accreted by the
central protostar, yielding an average mass accretion rate of
$\sim 10^{-6}$ to $\sim 10^{-5}$ M$_\odot$/yr. A much smaller amount of
mass and momentum reaches the outermost shell of cells at 20 AU and
is effectively removed from the calculation: the disk mass piles up in this
shell and is assigned zero radial velocity. The inner and outer boundary
conditions are designed to absorb incident mass and momentum, rather
than to reflect mass and momentum back into the main grid.

 As in Boss (2003, 2004), the models treat radiative transfer in the
diffusion approximation, with no radiative losses or gains occurring in 
regions where the vertical optical depth $\tau$ drops below 10.
In very low density regions interior to the initial disk structure, 
the disk temperature is assumed to be the same as that of the initial
disk at that radial location. Above the disk, the temperature
is set equal to the envelope temperature (Table 1). The envelope
temperatures of 30K to 50K are conservatively high estimates for
M dwarf protostars, given that temperatures of $\sim$ 50K appear to be 
appropriate for solar-mass protostars (Chick \& Cassen 1997).
 
\section{Initial Conditions}

 The models calculate the evolution of a 0.1 or $0.5 M_\odot$ central 
protostar surrounded by a protoplanetary disk with a mass ($M_d$) ranging 
from 0.021 to 0.065 $M_\odot$, respectively, between 4 AU and 20 AU 
(Table 1). The initial protoplanetary disk structure is based on
the following approximate vertical density distribution (Boss 1993) for
an adiabatic, self-gravitating disk of arbitrary thickness in
near-Keplerian rotation about a point mass $M_s$

$$ \rho(R,Z)^{\gamma-1} = \rho_o(R)^{\gamma-1} $$
$$ - \biggl( { \gamma - 1 \over \gamma } \biggr) \biggl[
\biggl( { 2 \pi G \sigma(R) \over K } \biggr) Z +
{ G M_s \over K } \biggl( { 1 \over R } - { 1 \over (R^2 + Z^2)^{1/2} }
\biggr ) \biggr], $$

\noindent where $R$ and $Z$ are cylindrical coordinates,
$\rho_o(R)$ is a specified midplane density, and $\sigma(R)$ is a specified
surface density. The disk surface occurs where $\rho(R,Z) = 0$.
The adiabatic pressure (used only for defining
the initial model -- the radiative transfer solution includes a full
thermodynamical treatment) is defined by $p = K \rho^\gamma$,
where $K$ is the adiabatic constant and $\gamma$ is the adiabatic
exponent. The adiabatic constant is $K = 1.7 \times 10^{17}$ (cgs units)
and $\gamma = 5/3$ for the initial model. The radial variation of 
the midplane density is chosen to be a power law that ensures
near-Keplerian rotation throughout the disk

$$\rho_o(R) = \rho_{o4} \biggl( {R_4 \over R} \biggr)^{3/2}, $$

\noindent where $\rho_{o4}$ varies between $1.3 \times 10^{-11}$
g cm$^{-3}$ and $6.0 \times 10^{-11}$ g cm$^{-3}$ (Table 1) and
$R_4 = 4$ AU. The resulting disk surface density falls off roughly
as $\sigma \propto r^{-1}$ over most of the disk, falling to
$\sigma \propto r^{-3/2}$ near the outer edge. A low density halo 
$\rho_h$ of gas and dust infalls onto the disk, with

$$ \rho_h(r) = \rho_{h4} \biggl ( {R_4 \over r} \biggr)^{3/2}, $$

\noindent where $\rho_{h4} = 1.0 \times 10^{-14}$ g cm$^{-3}$
and $r$ is the spherical coordinate radius. 

 The initial velocity field vanishes inside the disk, except for Keplerian
rotation, while in the halo the initial velocity field is given (based on
conservation of energy) by

$$ v_r = - \biggl( {G M_s \over r } \biggr)^{1/2} \cos \theta, $$

$$ v_\theta = \biggl( {G M_s \over r } \biggr)^{1/2} \sin \theta, $$

$$ v_\phi = \biggl( {G M_s \over r } \biggr)^{1/2}. $$

\noindent The translational ($v_r, v_\theta$) velocity field in the halo
is simply vertical infall toward the disk midplane, and the azimuthal
velocity is taken to be Keplerian. The chosen velocity field is
an analytical approximation that is convenient to implement and that
retains the essence of more exact solutions (e.g., Cassen \& Moosman
1981).
   
 The initial disk temperatures are based on the models of Boss (1995),
with midplane temperatures of 300 K at 4 AU, decreasing monotonically
outward to a distance of $\sim 6.7$ AU, where they are assumed to
become uniform at an outer disk temperature of $T_o$. $T_o$ is
varied in order to make sure that the outer disk is marginally
gravitationally unstable in terms of the gravitational stability
parameter $Q$, i.e., initial minimum values of $Q \sim 1.5$ are
assumed (Table 1). In low optical depth regions such as the disk envelope, 
the temperature is assumed to be equal to a fixed value, $T_e$,
with $T_e$ varied between 30 K and 50 K, and chosen to be
similar to $T_o$. The Rosseland mean opacities used in the radiative 
transfer solution have been updated to include the dust grain 
opacities calculated by Pollack et al. (1994).

\section{Results}

 Table 1 lists the initial conditions for the models with
$N_\phi = 256$. The models with $N_\phi = 512$ were started
at a point in the evolution of the $N_\phi = 256$ models 
when the disks were forming spiral arms and clumps dense enough
that the Jeans conditions were on the verge of being violated.
At that time, the $N_\phi = 256$ models were doubled
in azimuthal resolution to provide starting conditions
for the $N_\phi = 512$ models. 

 Figure 1 depicts the initial surface density profile for model 5C,
showing that the inner regions begin from a state of gravitational
stability, while the outer regions are closer to
marginal gravitational stability ($Q_{min} = 1.50$
for model 5C). The increased gravitational stability inside
$\sim 6$ AU is a direct result of the assumed higher initial
disk temperatures inside this radius (Boss 1995). Figure 1 also
illustrates that the assumed initial disk model does not result
in a single power law for $\sigma(r)$, but rather a smooth
falloff with radius toward $\sigma \propto r^{-3/2}$ near the
20 AU outer boundary.

 Figure 2 shows that model 5CH has formed a number of clearly
defined clumps after just 208 years of evolution around 
a protostar with a mass of $0.5 M_\odot$. Clump formation occurs 
soon after the initial disk develops spiral arms that transport 
angular momentum outward and mass inward, resulting in a 
significant depletion of the gas from the innermost disk 
(Figure 3). This occurs in spite of the inner region being gravitationally
stable -- gravitational torques from one-armed spirals propagating
at $\sim 8$ AU and beyond are able to drive disk gas inward and
onto the central protostar. Several Jupiter masses of gas are
accreted from the inner disk by the central protostar.

 Strong spiral arms begin forming first in the innermost, unstable
region of the disk, i.e., around $\sim$ 6 AU, because of the combination
of relatively low $Q$ and short orbital periods there. Spiral 
arms form at greater distances as the evolution proceeds.
The densest clumps tend to form at distances of $\sim$ 8 AU for the
same reasons, which are ultimately tied to the initial temperature 
profiles in the disk midplanes (Figure 1; Boss 1995).

 The densest clump seen in Figure 2 has a mass of $0.93 M_{J}$
within 0.01 of the maximum clump density of 
$8.8 \times 10^{-10}$ g cm$^{-3}$ (Figure 4). This mass is larger than
the Jeans mass at the average density ($2.3 \times 10^{-10}$ g cm$^{-3}$)  
and temperature (69 K; Figure 5) of the clump of $0.69 M_J$, 
implying that the clump is gravitationally bound in the absence of 
shear. Note that while
the average clump temperature is 69 K, peak clump temperatures
are $\sim$ 100 K, double the minimum outer disk temperature
($T_o$) in this model of 50 K -- the highest temperatures tend to
occur near the edges of the clump where the disk gas is being 
compressionally heated. The ratio of thermal
to gravitational energy for this clump is 0.80, also implying
that it is gravitationally bound. 

 The effective spherical radius of the densest clump in Figure 2
is 0.82 AU, which is somewhat larger than the critical
radius for tidal stability (Boss 1998) for this clump of 0.68 AU.
However, given that the clump is banana-shaped at this point in
time, with its minimum extent laying along the critical radial
direction, the clump should be capable of contracting to smaller
sizes and higher densities and should thus survive, though
perhaps not before having lost some mass through tidal forces.
Mass is also likely to be gained by accretion from the disk
once the clump becomes a well-defined protoplanet, as was found
in the virtual protoplanet models of Boss (2005). The clump
has a nearly circular orbit at this point, with an eccentricity
of 0.0005 and semimajor axis of 7.7 AU. If the clump survives
to form a protoplanet and undergoes substantial inward migration
to a stable orbit, its final orbital eccentricity is likely to be
unrelated to this initial eccentricity, but will depend on the 
mechanism responsible for the inward migration.

 Figure 6 illustrates that the densest clump seen in Figure 2 is 
well-resolved with respect to the Jeans criteria. Hence, this clump, 
as well as the somewhat less dense clumps seen in Figure 2, do not 
appear to be spurious results. Once the clumps form, they contract
to higher densities and may then begin to violate the Jeans
criterion for the radial coordinate, the most restrictive
Jeans criterion at the orbital distance of the clumps
(Figure 6). However, given that the clumps formed without
violating the Jeans criteria, their subsequent contraction
to higher densities must be viewed as physically realistic,
though any indication of sub-fragmentation during such a 
poorly-resolved phase would be suspicious and is in fact not seen.
The clumps attain masses of $\sim 1.1$ to $1.3 M_J$ within
10 yrs after the phase shown in Figure 2, with effective 
spherical radii less than or comparable to their critical 
tidal radii. Their orbital eccentricities increase to
$\sim 0.1$ during this time interval. 

 The clumps seen in model 5CH in Figure 2 eventually are torn apart 
by tidal forces as the present grid-based code is unable to provide
the locally higher spatial resolution needed for the clumps to
continue their contraction to ever higher and higher densities.
The models of Boss (2005) showed that clumps become better
defined and longer-lived as the spatial resolution is
increased, but the question of clump survival and evolution
will ultimately require treatment by an adaptive mesh
refinement (AMR) code. Smoothed-particle-hydrodynamics (SPH)
models of disk instabilities with locally defined smoothing
lengths (Mayer et al. 2002, 2004) have already demonstrated the 
need for such locally enhanced spatial resolution. Even though
the clumps seen in Figure 2 are doomed to disappear eventually
by the nature of the present calculations, new clumps continue to
form and evolve, implying that the disks are determined to
eventually form self-gravitating clumps.
 
 All of the other models of disks orbiting $0.5 M_\odot$ stars 
(models A, AH, B, BH, and D) behave in the same manner as
models C and CH: the disks develop strong spiral arms that
interact and form self-gravitating clumps with masses
ranging from $\sim 0.3$ to $1.9 M_J$. These masses appear
to be roughly consistent with the minimum masses of the gas giant
planets orbiting the M dwarf GJ 876 (Marcy et al. 1998, 2001).

 We now turn to the models with disks orbiting $0.1 M_\odot$ 
stars (models 1A, 1B, 1C, and 1D). Figure 7 shows the outcome
of 446 yrs of evolution of model 1C, which had a disk with
an initial mass of $0.021 M_\odot$. Even with these low values of
the stellar and disk masses, model 1C managed to form a
self-gravitating clump in a very short period of time.

 The densest clump seen in Figure 7 at 4 o'clock for model 1C 
has a mass of $1.1 M_J$ within 0.05 of the maximum clump density of 
$1.6 \times 10^{-10}$ g cm$^{-3}$. This mass is larger than
the Jeans mass at the average density ($2.2 \times 10^{-11}$ g cm$^{-3}$)  
and temperature (39 K; Figure 8) of the clump of $0.96 M_J$, 
implying that the clump once again is gravitationally bound. 
Figure 8 again shows that the highest temperatures do not
necessarily occur at the location of the density maxima,
but rather near the clump edges. The clump has an orbital
eccentricity of 0.12 and a semimajor axis of 7.4 AU at this time.

 Models 1A and 1B also formed gravitationally bound clumps.
Model 1D formed clumps, but the densest clump analyzed did not 
appear to be quite dense enough to be gravitationally bound, 
at least for the limited number of time steps when the 
calculational data was saved for later analysis. Periodic clump 
formation persisted throughout the 1300 yrs that 
model 1D was evolved, so there is no clear evidence that
the disk instability process could not result in the formation
of planetary-mass companions to even lower mass objects,
i.e., brown dwarfs, provided that these objects have marginally
gravitationally unstable disks at some phase of their evolution.

 Movies of models 1A, 1B, 1C, 1D, 5A, 5AH, 5B, and 5BH in mpeg
format may be downloaded from the following web site: 
http://www.dtm.ciw.edu/boss/ftp/mpeg/mdwarf/.

\section{Conclusions}

 These models suggest that there is no reason why M dwarf stars should 
not be able to form gas giant protoplanets rapidly, if a disk 
instability can occur. This prediction differs fundamentally from that
of the core accretion mechanism, which does not appear to be
able to form gas giant planets around M dwarf on time scales
shorter than typical disk lifetimes (Lauglin et al. 2004).
Disk instability may even be able to form planetary-mass objects
in orbit around brown dwarfs with suitably unstable disks, though
this has not been demonstrated by the present set of models.

 Ongoing and future extrasolar planet searches will answer the question
of whether or not gas giant planets form frequently around low mass stars. 
If M dwarfs turn out not to have very many gas giant planets, this
could be interpreted as either a failure of disk instability to
produce gas giant planets, or as a failure of M dwarfs to have
disks massive enough to be marginally gravitationally unstable.
If M dwarfs do turn out to have a significant frequency of gas
giant planets, this would imply that disk instability is responsible
for their formation, given the apparent inability of core accretion
to form gas giants fast enough (Laughlin et al. 2004). 

 Regardless of the situation for the gas giant planets, the 
Neptune-mass planets that have been found around M dwarfs 
are unlikely to have formed by disk instability, unless massive
gaseous envelopes can be removed by stellar heating and irradiation, 
a process that may be reasonably efficient for solar-mass stars
(Vidal-Madjar et al. 2004; Hebrard et al. 2005). Rather, such planets 
are likely to have formed by the same collisional accumulation process
that led to the terrestrial planets in our Solar System. Such a finding 
would not preclude the existence of gas giant planets orbiting at
greater distances from M dwarfs, as is the case for GJ 876,
planets which presumably formed by disk instability.

 Given the extreme imbalance in the numbers of the closest stars of 
different spectral types,  
M dwarfs are a natural choice for astrometric planet searches, where 
closeness is a primary virtue. The low mass of the primary helps as
well to enable astrometric detections of even lower mass planets. 
Ground-based efforts are underway at Carnegie's Las Campanas Observatory
to detect brown dwarf and gas giant planet companions to M, L, and
T dwarfs by astrometry (Boss et al. 2006). Space-based astrometric 
searches by NASA's {\it Space Interferometry Mission (SIM)} should
lower the detectable planet mass to Earth-masses or below
around M dwarfs. Space-based transit missions (COROT, Kepler) will 
be able to detect many giant planets orbiting M dwarfs, as well as
the ongoing ground-based microlensing and spectroscopic searches.
We can expect that all of these efforts will combine to determine the 
gas giant planet frequency around M dwarfs, and thereby help to 
determine the most likely mechanism for the formation of their
giant planets.

 I thank Fred Adams for his valuable comments on the manuscipt and
Sandy Keiser for her critical computer systems expertise. This 
research was supported in part by NASA Planetary Geology and Geophysics
grant NNG05GH30G and by NASA Astrobiology Institute grant NCC2-1056.
The calculations were performed on the Carnegie Alpha Cluster, 
the purchase of which was partially supported by NSF Major Research
Instrumentation grant MRI-9976645.

\begin{figure}
\vspace{-2.0in}
\plotone{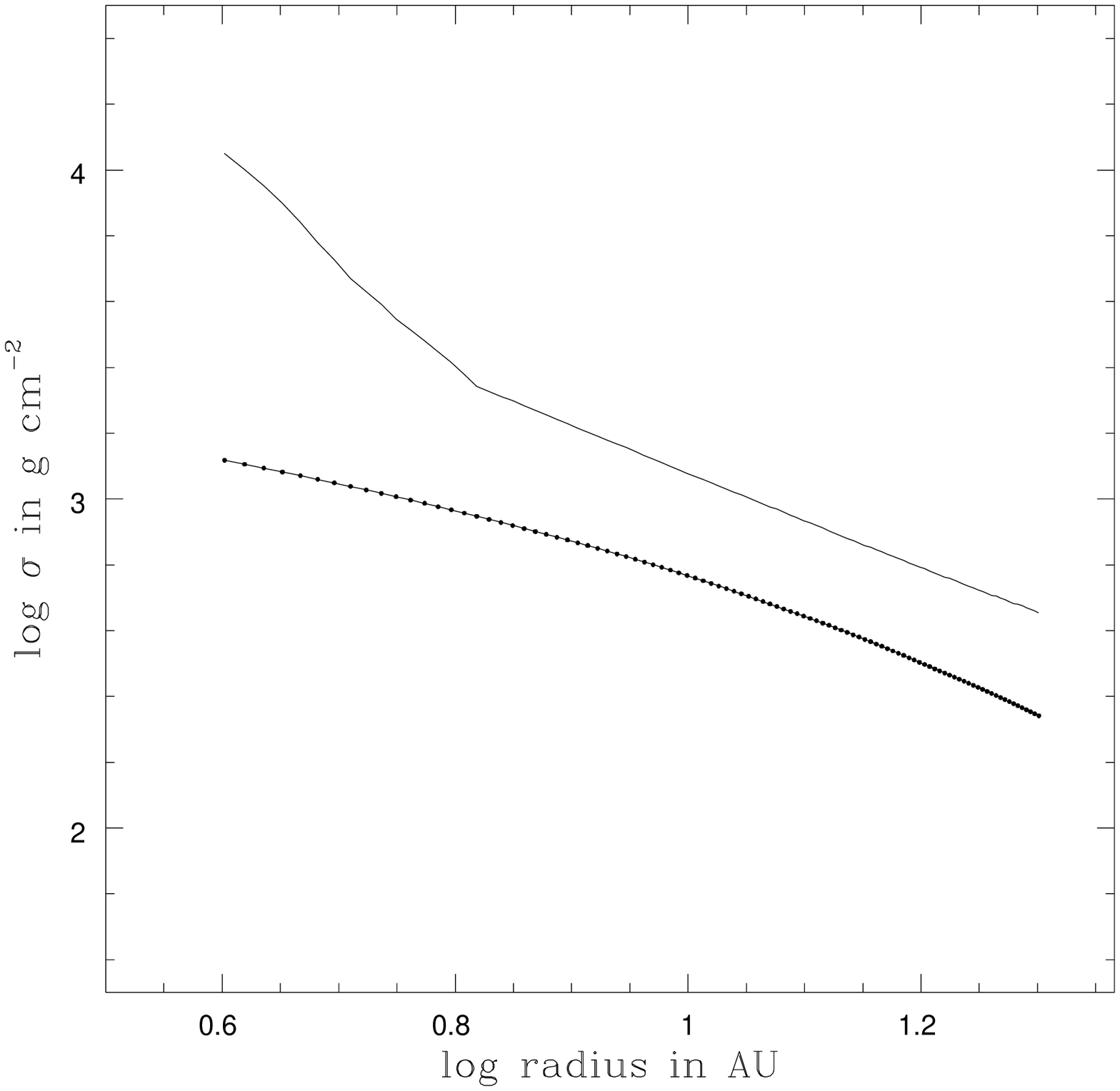}
\caption{Radial (azimuthally averaged) profile of the 
disk gas surface density (dots) in the initial model 5C, 
compared to the surface density needed for $Q = 1$ (solid line).
The initial disk surface density is too low by about a factor of two 
for the outer disk to have $Q = 1$.}
\end{figure}

\begin{figure}
\vspace{-2.0in}
\plotone{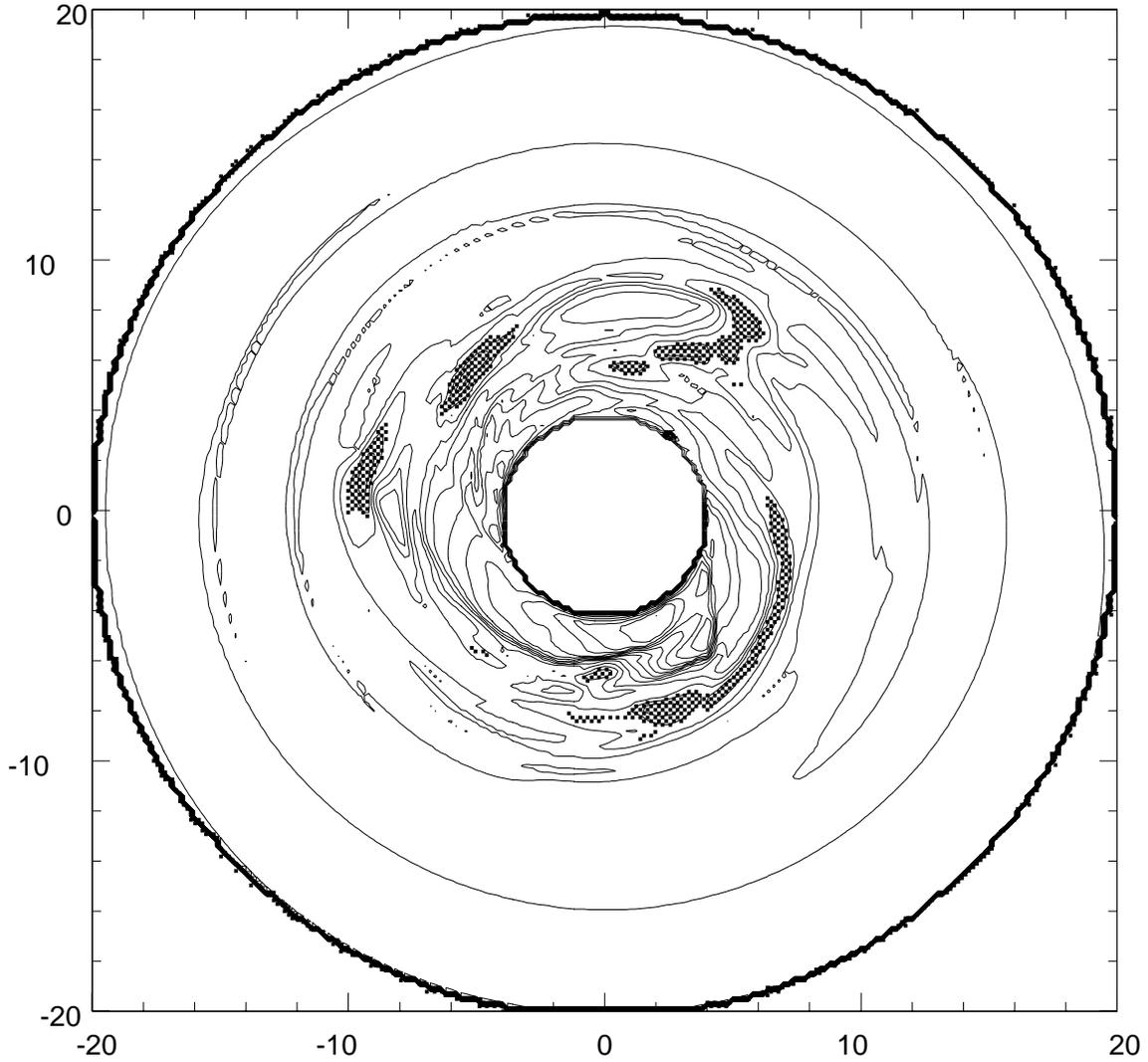}
\caption{Equatorial density contours for model 5CH after 
208 yrs of evolution. The entire disk 
is shown, with an outer radius of 20 AU and an inner radius of 4 AU, 
through which mass accretes onto the central protostar. Hashed regions 
denote spiral arms and clumps with densities higher than $10^{-10}$ 
g cm$^{-3}$. Density contours represent factors of two change in density.}
\end{figure}

\suppressfloats

\begin{figure}
\vspace{-2.0in}
\plotone{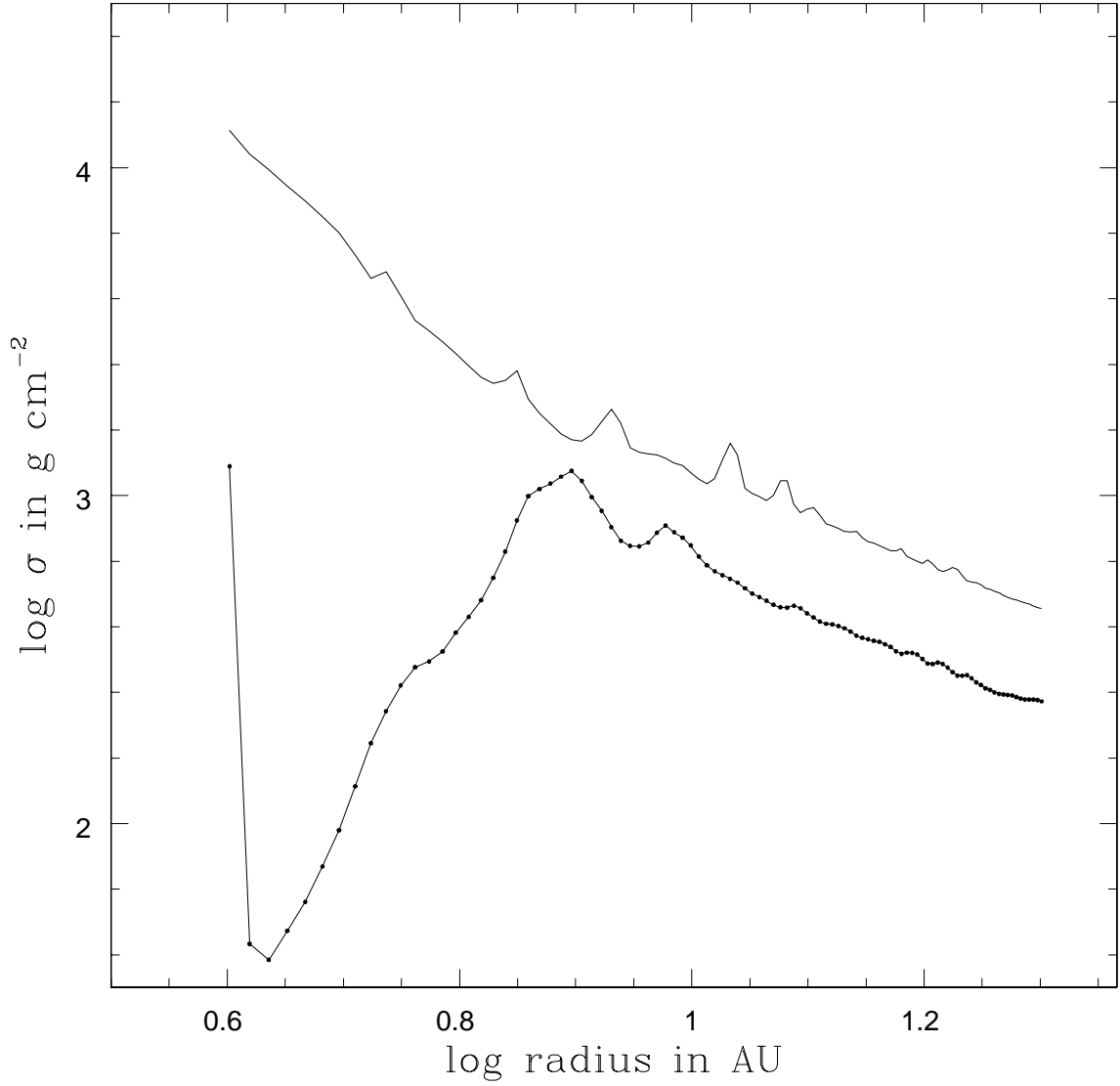}
\caption{Radial (azimuthally averaged) profile of the disk gas 
surface density (dots) for model 5CH after 208 yrs of evolution,
compared to the surface density needed for $Q = 1$ (solid line).
The innermost disk gas (several Jupiter-masses) has largely 
accreted onto the central protostar.}
\end{figure}

\begin{figure}
\vspace{-2.0in}
\plotone{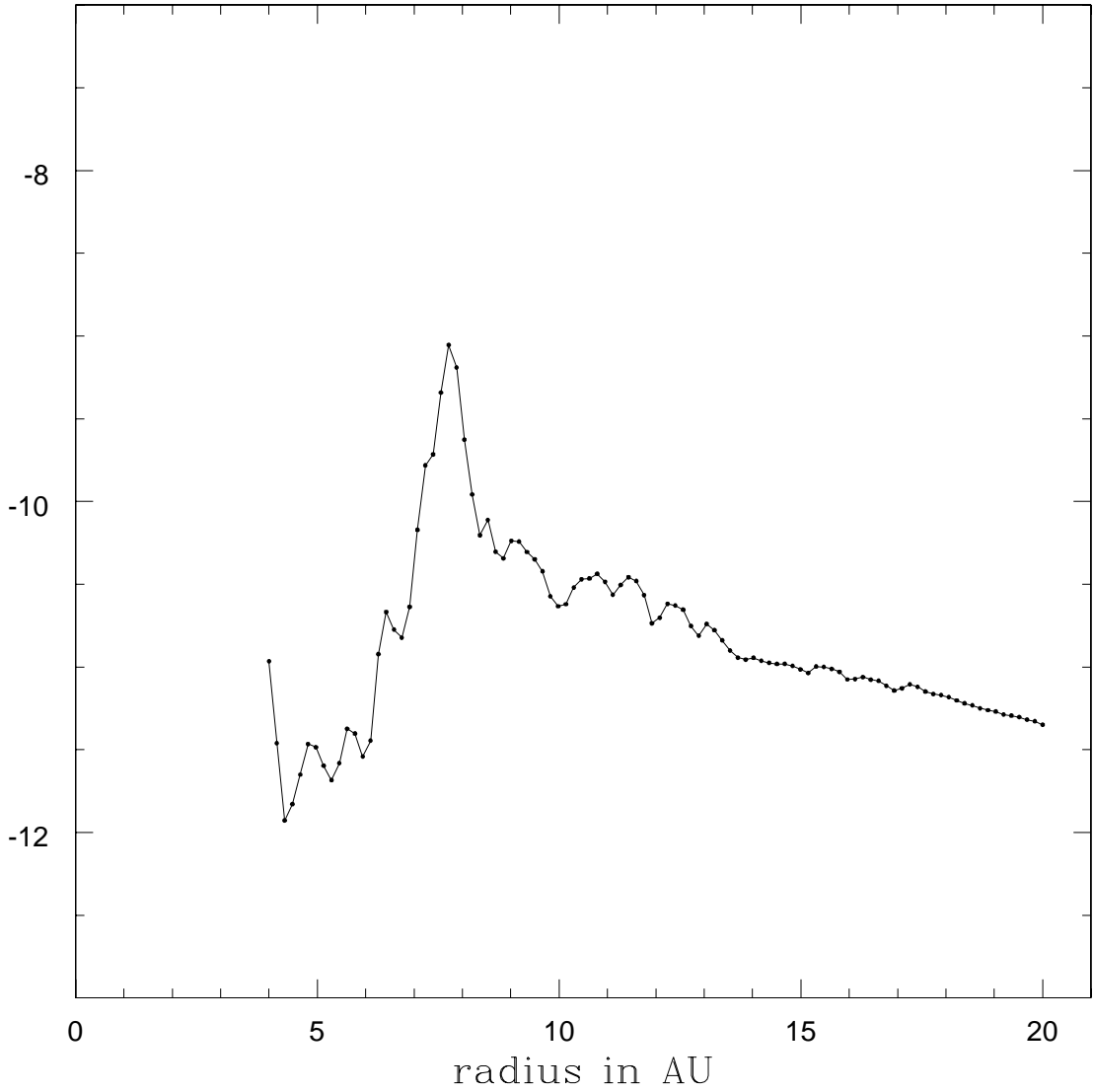}
\caption{Radial density profile (log of the density in g cm$^{-3}$) 
passing through the densest clump (with a mass of $\sim 0.93 M_J$) 
seen in Fig. 2 at 9 o'clock for model 5CH.}
\end{figure}

\begin{figure}
\vspace{-2.0in}
\plotone{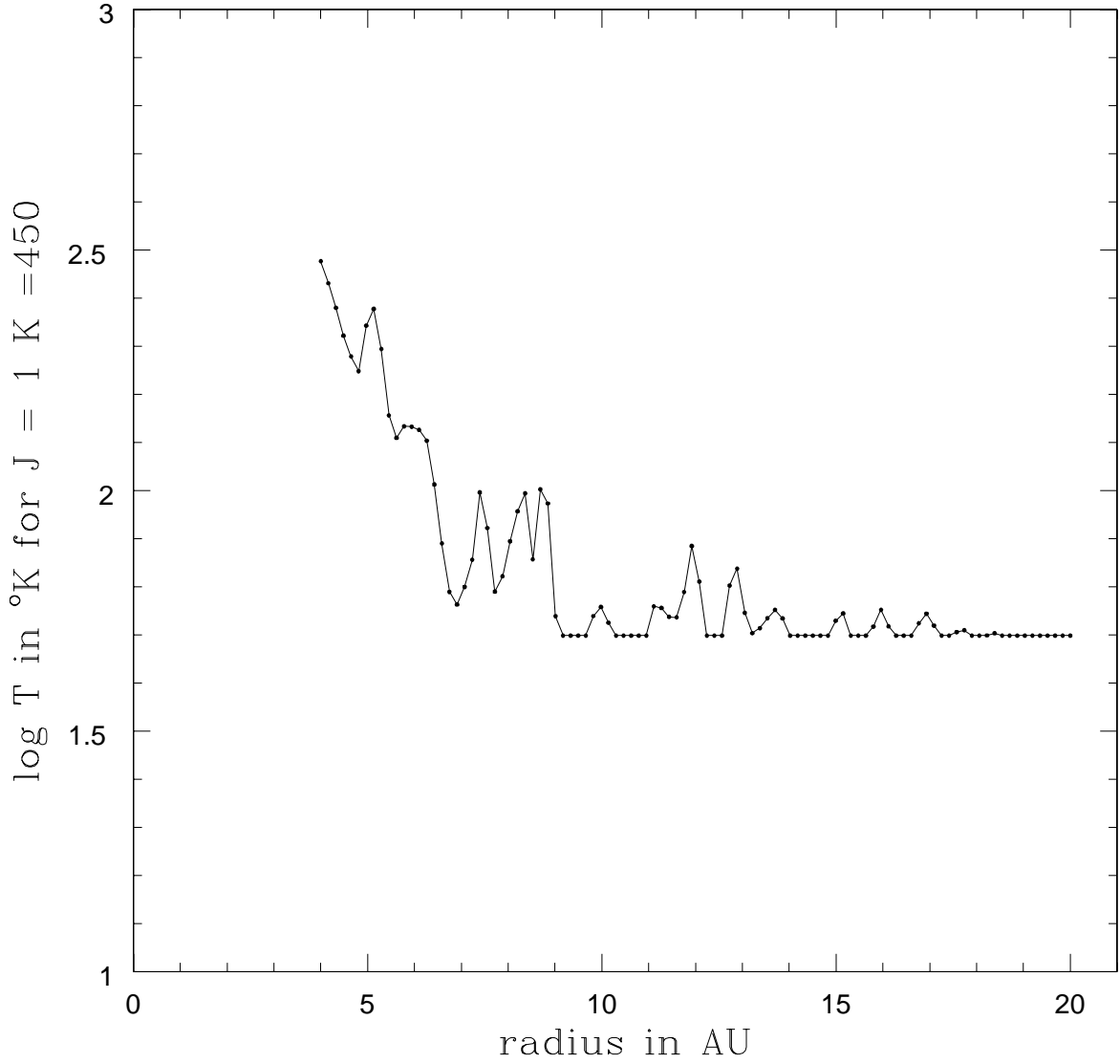}
\caption{Radial temperature profile for model 5CH, as in Fig. 4, taken
along a ray through the densest clump in Fig. 2. Temperatures in
the outer disk are not allowed to drop below their initial values,
which accounts for the thermal floor evident outside 9 AU.}
\end{figure}

\begin{figure}
\vspace{-2.0in}
\plotone{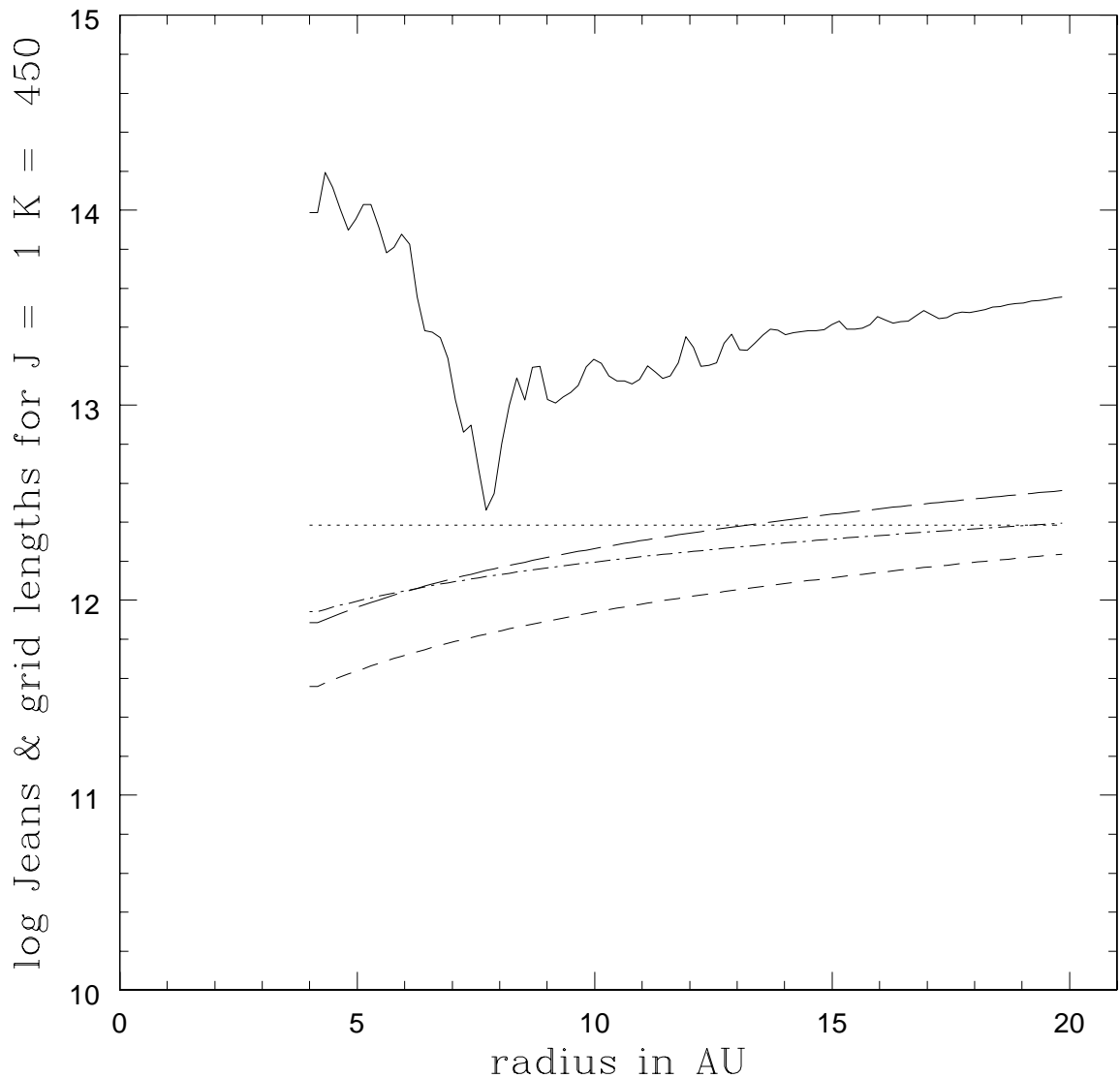}
\caption{Critical Jeans length criterion (solid line) compared to the four
spherical coordinate grid spacings ($\Delta x_r = \Delta r$ = dotted,
$\Delta x_\theta = r \Delta \theta$ =  dashed,
$\Delta x_\phi = r sin \theta \Delta \phi$ = long-dashed,
and $\Delta x$ = dot-dashed) for model 5CH after 208 yrs. 
The radial profile denotes the values for a ray that passes
through the density maximum seen at 9 o'clock in Figure 2,
leading to the dip in the Jeans criterion at $\sim$ 8 AU.}
\end{figure}

\suppressfloats

\begin{figure}
\vspace{-2.0in}
\plotone{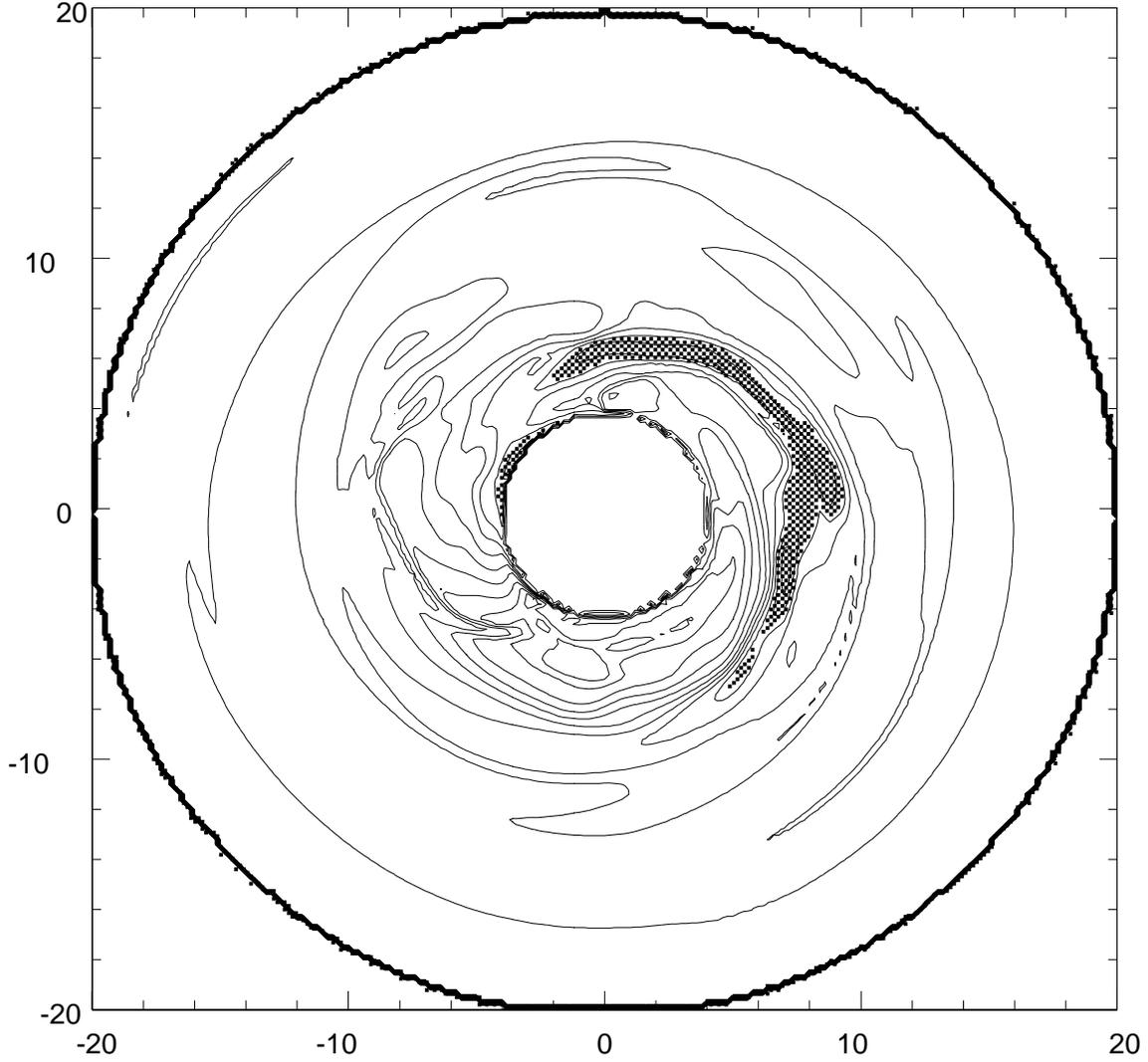}
\caption{Equatorial density contours for model 1C after 446 yrs of 
evolution, as in Fig. 2. In a Fourier decomposition of the midplane
density in this model, the $m = 1$ mode dominates with a maximum
amplitude $a_{m=1} \approx 1.5$ at $\sim$ 8 AU, where 
$a_{m=2} \approx 0.8$ and $a_{m=3} \approx 0.2$.}
\end{figure}

\begin{figure}
\vspace{-2.0in}
\plotone{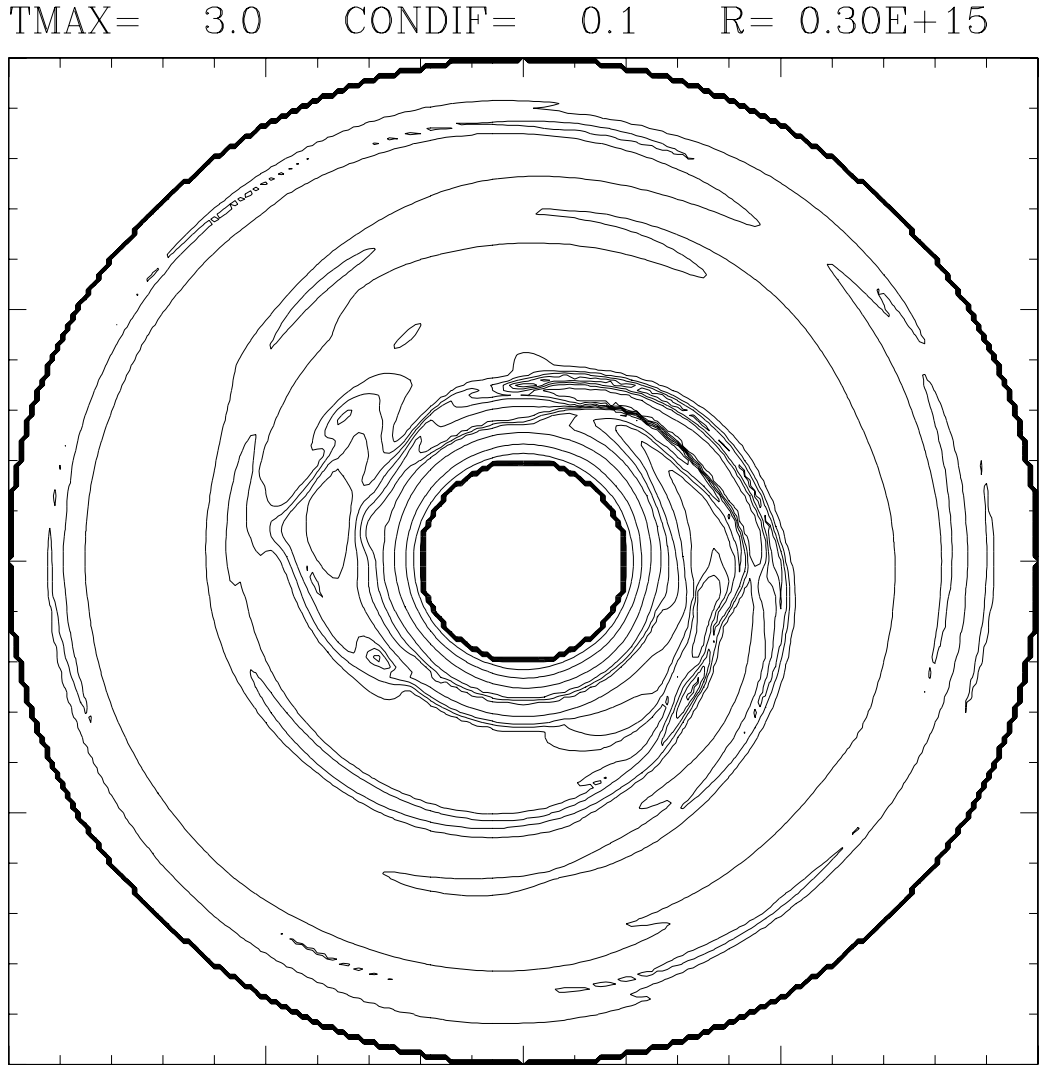}
\caption{Equatorial temperature contours for model 1C after 446 yrs of 
evolution, as in Fig. 7, except that contours represent factors of 
1.3 change in temperature.}
\end{figure}

\clearpage
\begin{deluxetable}{ccccccccc}
\tablecaption{Initial conditions for the models.\label{tbl-1}}

\tablehead{\colhead{model} & 
\colhead{$M_s/M_\odot$ } & 
\colhead{$M_d/M_\odot$ } & 
\colhead{$\rho_{o4}$ (g cm$^{-3}$) } & 
\colhead{$T_o$ (K) } & 
\colhead{$T_e$ (K) } & 
\colhead{$min(Q_i)$ } & 
\colhead{$N_\phi$ } & 
\colhead{$N_{Ylm}$ } }

\startdata

1A & 0.1 & 0.031 & $1.9 \times 10^{-11}$ & 50K & 50K & 1.44 & 256 & 32 \\

1B & 0.1 & 0.031 & $1.9 \times 10^{-11}$ & 60K & 50K & 1.60 & 256 & 32 \\

1C & 0.1 & 0.021 & $1.3 \times 10^{-11}$ & 30K & 35K & 1.53 & 256 & 32 \\

1D & 0.1 & 0.021 & $1.3 \times 10^{-11}$ & 35K & 35K & 1.65 & 256 & 32 \\

5A  & 0.5 & 0.041 & $3.9 \times 10^{-11}$ & 20K & 30K & 1.40 & 256 & 32 \\

5AH & 0.5 & 0.041 & $3.9 \times 10^{-11}$ & 20K & 30K & 1.40 & 512 & 48 \\

5B  & 0.5 & 0.041 & $3.9 \times 10^{-11}$ & 25K & 30K & 1.56 & 256 & 32 \\

5BH & 0.5 & 0.041 & $3.9 \times 10^{-11}$ & 25K & 30K & 1.56 & 512 & 48 \\

5C & 0.5 & 0.065 & $6.0 \times 10^{-11}$ & 50K & 50K & 1.50 & 256 & 32 \\

5CH & 0.5 & 0.065 & $6.0 \times 10^{-11}$ & 50K & 50K & 1.50 & 512 & 48 \\

5D & 0.5 & 0.065 & $6.0 \times 10^{-11}$ & 60K & 50K & 1.63 & 256 & 32 \\

\enddata
\end{deluxetable}
\clearpage

\suppressfloats

\end{document}